\documentstyle[aps,preprint]{revtex}

\newcommand{\SdH}{Shubnikov-de Haas }

\begin{document}
\title{Large transconductance oscillations in a single-well vertical
Aharonov-Bohm interferometer}
\author{Vincenzo Piazza$^1$, Fabio Beltram$^1$, Werner Wegscheider$^{2,3}$,
Chi-Te Liang$^{4,5}$, and M. Pepper$^4$}
\address{$^1$ Scuola Normale Superiore and Istituto Nazionale per la
Fisica della Materia, I-56126 Pisa, Italy}
\address{$^2$ Walter Schottky Institute, Munich, Germany}
\address{$^3$ Universit\"at Regensburg, Regensburg, Germany}
\address{$^4$ Cavendish Laboratories, University of Cambridge, 
Cambridge CB3 0HE, United Kingdom}
\address{$^5$ Department of Physics, National Taiwan University, Taipei 106,
Taiwan}
\maketitle
\begin{abstract}
Aharonov-Bohm (AB) interference is reported for the first time
in the conductance of a vertical nanostructure based on a
{\em single} GaAs/AlGaAs quantum well (QW).
The two lowest subbands of the well are spatially separated by
the Hartree barrier originating from electronic repulsion in
the modulation-doped QW and
provide AB two-path geometry. 
Split-gates control the in-plane electronic momentum dispersion.
In our system, we have clearly demonstrated AB interference in
both electrostatic and magnetic modes. In
the latter case the magnetic field was applied parallel to the QW
plane, and perpendicular to the $0.02 \, \mu \rm{m^2}$ AB loop. 
In the electrostatic mode of operation the single-QW
scheme adopted led to large transconductance oscillations with relative 
amplitudes exceeding 30\%. The 
relevance of the present design strategy for the implementation
of coherent nanoelectronic devices is underlined.
\end{abstract}
\newpage
The pioneering work by Aharonov and Bohm had far-reaching fundamental
implications as it pointed out the significance of
potentials with respect to forces in quantum-mechanics\cite{aharonov}.
In their article, Aharonov and Bohm demonstrated that
electrons propagating in the two arms of a ring-like structure
experience a relative phase shift of 
$e / \hbar \int{\bf A}\cdot d{\bf l}$ (where ${\bf A}$ is the
magnetic vector potential and the integral is performed
along the ring) when a magnetic flux encircled
by the ring is applied. This is true even if the classical Lorentz force is
exactly zero along the electron path. 
A similar effect was predicted if one of the 
two arms was subjected to a change of its electrostatic potential, while
keeping the electric field at zero by means of Faraday cages.
If electrons injected and 
collected into/from the structure by means of leads propagate
coherently, the phase shift was shown to manifest macroscopically
in observable quantities such as conductivity. The latter was
predicted to oscillate following
cycles of constructive and destructive electronic interference.

Since its proposal, Aharonov-Bohm (AB) interference was 
demonstrated in several 
mesoscopic systems\cite{sharvin,datta1,timp,ford,wharam,okuda2}. 
The unprecedented
advancement in semiconductor growth and nanofabrication technologies has
opened the way to device applications of this effect. Indeed
semiconductor devices based on electrostatic AB effect were proposed as 
building blocks for the development of coherent 
nanoelectronics\cite{datta2,okuda1}. 
The advantages offered by quantum-interference devices with 
respect to conventional ``classical'' ones are very significant:
a quantum interference transistor requires much smaller gate
voltages compared to conventional FETs (some 
milliVolts compared to Volts) and this is expected to lead to higher
transconductance and extremely small power dissipation \cite{datta3}.

AB effect relies on electronic phase coherence, however, and its
observation in semiconductor
heterostructures with marked and {\em useful} conductance oscillations demands 
high-quality materials and very careful design.
Datta {\em et al.}\cite{datta1} pioneered the field and
demonstrated for the first time magnetic AB oscillation in a GaAs/AlGaAs
heterostructure with a contrast (defined as
the ratio between the peak-to-peak amplitude and the mean value) 
smaller than 1\%. Since then significant 
progresses were made in the quality of materials and in 
structure design, but even some more recent demonstrations of
AB effect in semiconductor ring structures showed 
a contrast of the magnetic AB signal of only about 10\% \cite{appenzeller}.

In light of successful applications of the AB effect to quantum
devices, it is of crucial importance to achieve much higher contrasts
and implement the electrostatic AB configuration. 
In this communication we propose and demonstrate a novel 
scheme for an AB interferometer based on a 
single modulation-doped wide GaAs quantum well. The two transport 
paths originate from the two lowest subbands of the well and 
take advantage of the 
$\sim \rm{\AA}$ resolution achieved by epitaxial growth techniques. 
Spatial separation between the two electronic paths 
is given by the formation of a Hartree barrier in the well, 
lateral confinement is provided by surface gating. 

The samples were fabricated from a high-mobility single-quantum-well 
GaAs/AlGaAs heterostructure containing a two-dimensional electron gas (2DEG)
with a Hall mobility of 
$5 \cdot 10^{6} \,{\rm cm}^2 {\rm V}^{-1} {\rm s}^{-1}$ and a 
total carrier density
of $5.8 \cdot 10^{11} \,{\rm cm}^{-2}$ at 4.2 K after illumination.
Carriers were distributed in
the two lowest subbands and \SdH measurements yielded 
$3.9 \cdot 10^{11} \,{\rm cm}^{-2}$ electrons in the ground
subband and $1.7 \cdot 10^{11} \,{\rm cm}^{-2}$ in the first-excited one. 
The 60 nm GaAs quantum well is located 75 
nm below the surface and is embedded between 
Al$_{0.25}$Ga$_{0.75}$As barriers. Two $\delta$-doping layers are placed 
35 nm (surface side) and 60 nm (substrate side) from the well.
The choice of a wide 
single well instead of a double-quantum-well structure was motivated by the 
twofold need for spatially-separated transport paths and 
long coherence lengths. In this scheme, spatial separation is guaranteed 
by the Hartree barrier formed by Coulomb repulsion between carriers
populating the modulation-doped quantum well (see inset in Fig.~1).
At the same time, the 
absence of a central AlGaAs barrier reduces interface roughness 
scattering and alloy scattering.

One-dimensional leads are necessary to avoid smearing of
the interference pattern in the electrostatic AB effect \cite{datta2}. Indeed,
while for the magnetic AB effect phase shifts depend on the
area enclosed by the two paths, for the electrostatic AB effect
the shift depends on
the transit time of electrons in the two arms of the ring \cite{datta2}.
Consequently injection of an electron beam with no in-plane 
momentum dispersion is required for AB observation.
To this end, an injector and a collector were realised on a Hall bar
by patterning two
long (5 ${\rm \mu m}$) split gates on the surface of the sample. 
(C, D and E, F in Fig.~1). The geometrical width of the gap between 
the electrodes is 0.5 ${\rm \mu m}$.) 
By appropriately biasing these gates it is possible to
tune the number of modes propagating in the leads and, in the limit of
zero temperature, in-plane momentum dispersion 
will be suppressed when a single mode is available for transport. 

Given an estimated coherence length of about 1 ${\rm \mu m}$, 
electronic transport in the leads is expected to be diffusive.
This is required in our scheme to avoid coherent reflections
at their ends that could hinder the observation 
of the main interference effect.
The injector and collector were individually characterised by 
measuring the
low-temperature conductance of the sample versus the split-gate 
biases $V_{\rm CD}$ and $V_{\rm EF}$\cite{exp}. 
None of them displayed clear signatures of conductance quantisation, 
confirming the diffusive nature of electron transport.

A third split gate (A, B) was used to define the ballistic one-dimensional 
region and tune the phase difference between propagating paths. 
Electrons can be reflected back or transmitted to the collector
region depending on the interference being destructive or constructive.
The split-gate length was chosen of 0.6 ${\rm \mu m}$,
shorter than the estimated coherence length.
Indeed the conductance of the device versus the bias 
($V_{\rm AB}$) of the quantum point 
contact (QPC) defined by A and B displayed clearly quantised
steps.

Fig.~2 shows the conductance of the device at different values of 
$V_{\rm AB}$ as a function of $V_{\rm CDEF} = V_{\rm CD} = V_{\rm EF}$. 
Sweeping $V_{\rm CDEF}$ 
from zero bias toward negative values, a one-dimensional channel 
is created as the 2DEG below the gates is depleted and then 
progressively narrowed until channel pinch-off 
at $\approx -1$ V. The value of the conductance
in units of $2e^2/h$ at a fixed $V_{\rm AB}$ 
and at $V_{\rm CDEF} = 0$ can be read as the
number of propagating modes in the QPC.
For $-3 {\,\rm V} < V_{\rm AB} < -1 {\,\rm V}$, corresponding to 2 
to more than 3 modes in the QPC, very clear oscillations can be 
observed over a monotonically decreasing background as 
$V_{\rm CDEF}$ is swept below $\approx -0.5 {\,\rm V}$.
No oscillations could be seen when less 
than two modes are propagating 
in the QPC ($V_{\rm AB} < -3 {\,\rm V}$), that is when 
only a single path is available for transport.

Fig.~3 reports a set of oscillations obtained on one such 
device. Subtracting the contribution 
to the resistance due to the one-dimensional diffusive leads 
(${\rm \sim 10 k\Omega}$)
the contrast of the oscillations exceeds
30 \%. To our knowledge, this value is the 
highest reported in the literature for 
semiconductor-heterostructure electrostatic AB rings and
makes the present fabrication strategy of interest for device
applications. Indeed the present structure does implement a possible 
quantum-interference transistor that yields a normalised transconductance
$dI_D / dV_G \times 1/I_D \, \agt 35 \, V^{-1}$. 

In order to understand the nature 
of the states (paths) involved in the interference processes
it is necessary to distinguish between different situations.
For sake of simplicity, we focus on the case 
of two propagating modes in the QPC.
This configuration can occur in two different ways:
i) only the first two transverse modes 
originating from the lower quantum-well subband are below the
Fermi level and all modes originating from the upper subband
are depleted; ii) electrons occupy the first transverse mode 
of each one of the two quantum-well subbands.
To understand which of these possibilities is of interest here, 
we measured the conductance of our device
at appropriately chosen gate biases as a
function of an in-plane magnetic field perpendicular 
to the current.  The magnetic field will couple to 
the loop formed by the two propagating paths 
and lead to magnetic AB oscillations
only when configuration ii) is realised.
As reported in Fig.~4, very clear conductance oscillations can be seen as 
a function of magnetic field. Consistently
with the results reported above, interference
effects occur only when $V_{\rm CDEF}$ is swept below
$\approx -0.5 \,{\rm V}$. Again, no oscillations can be resolved when 
a single mode is accessible in the QPC.
The period of AB oscillations is given by $\Delta B = h / e S$, where 
$S = L d \approx 0.02 \, \mu {\rm m}^2$ is
the area enclosed by the two paths, $d$ is the center-to-center distance
between the channels and $L$ is the length of the two channels.
For our sample we estimate $L \approx 0.6 \,{\rm \mu m}$
from the geometrical width of the QPC and $d \approx 35 \,{\rm nm}$, 
obtained from a self-consistent calculation of the two subband wavefunctions.
These values give a theoretical period $\Delta$B $\approx 0.19 \,{\rm T}$,
in good agreement with the values of $0.2 \,{\rm T}$ and $0.26 \,{\rm T}$
measured from the data with $V_{\rm CDEF} = -0.5 \,{\rm V}$ and 
$V_{\rm CDEF} = -1.2 \,{\rm V}$ respectively (in both cases 
$V_{\rm AB} = -2.5 \,{\rm V}$). The greater period measured
in the latter case is linked to the increased wavefunction 
confinement and consequent reduction of AB loop area.

In summary, we have demonstrated a novel scheme leading to
very high-contrast Aharonov Bohm interference
driven both by an electric and a magnetic field 
in a high-mobility heterostructure.
The analysis of the results allowed to unambiguously 
identify the states involved in the process. The present structure
does implement a quantum interference transistor device that in
its present configuration leads to a normalised
transconductance analogous to that of classical devices.

\vskip 0.5cm
This work was funded in part by the British Council and CRUI within the
British-Italian collaboration scheme. The work
at Cambridge was supported by the UK EPSRC. C.~T.~L. is grateful 
for the support from the NSC, Taiwan. 
Molecular beam epitaxial growth at the Walter Schottky Institut (TU Munchen)
is supported by the Deutsche Forschungsgemeinschaft via SFB 348 and
the Bundesministerium f\"ur Bildung und Forschung under contract 
01 BM 630/1. We would like to thank Spiros
Gardelis for his help in sample preparation.

\begin{figure}
\caption{Scanning electron micrograph of one of
the device studied. Cr-Au nanogates
are shown as bright regions and define a set of wires in the $x$ direction.
The Aharonov-Bohm ring is formed in the plane perpendicular to that of the
image. Path separation arises from the Hartree barrier formed by Coulomb 
repulsion between carriers populating the modulation-doped quantum well. 
The inset shows the calculated conduction-band
diagram (solid line) and the two lowest subband wavefunctions (dashed lines)
of the quantum well at equilibrium at 4.2K.} 
\end{figure}

\begin{figure}
\caption{Differential conductance measurements at 300 mK as a function of
the bias applied to gates C, D, E and F (see Fig.~1). Four curves are plotted
for different biases applied to gates A and B.}
\end{figure}

\begin{figure}
\caption{Differential resistance at 300 mK as a function of the bias
applied to gates C, D, E and F. Gates A and B were biased at -2.2 V.}
\end{figure}

\begin{figure}
\caption{Differential conductance at 300 mK as a function of the external
magnetic field applied in the $y$ direction (see Fig.1). The three panels
show data for different biases
applied to gates C, D, E and F. Gates A and B were biased at -2.5 V.}
\end{figure}


\begin{references}
\bibitem{exp} {All the measurements were performed, unless explicitly 
        stated, after illumination at 300 mK in a $^3$He cryostat by 
        four-terminal lock-in techniques, 
	with a 10 nA excitation current at 13 Hz.}
\end{references}
\end{document}